\documentclass[fleqn,usenatbib]{mnras}

\usepackage[T1]{fontenc}
\usepackage{ae,aecompl}

\usepackage{graphicx}	    
\usepackage{amsmath}	    
\usepackage[]{algorithm2e}  

\makeatletter
\def\@printed{
    \qquad\qquad\qquad Compiled using MNRAS \LaTeX\ style file v\@version}
\gdef\@journal{\hfill\@printed}
\def\bsp{}
\def\@oddfoot{}
 \def\@evenhead{}
 \def\@evenfoot{}
\def\ps@titlepage{\let\@mkboth\@gobbletwo
 \def\@oddhead{\footnotesize\@journal}
 \def\@oddfoot{\hfil}
 \def\@evenhead{\hfill}
 \def\@evenfoot{\hfil}
 \def\sectionmark##1{}
 \def\subsectionmark##1{}}
\def\abstract{\if@twocolumn
   \start@SFBbox\@abstract
 \else
   \@abstract
 \fi}
\def\endabstract{\if@twocolumn
   \endlist\finish@SFBbox
 \else
  \endlist
 \fi}
\def\@abstract{\list{}{%
    \listparindent\realparindent
    \itemindent\z@
    \labelwidth\z@ \labelsep\z@
    \leftmargin 1.5pc\rightmargin 1.5pc
    \parsep 0pt plus 2pt}\item[]%
    \reset@font\normalsize{\bf ABSTRACT}\\\reset@font\large
}
\def\@maketitle{\newpage
 \vspace*{7pt}
 {\raggedright \sloppy
  {\reset@font\huge \bf \@title \par}
  \vskip 23pt
  {\reset@font\LARGE
   \begin{tabular}[t]{@{}l@{}}\let\\=\author@nextline\@author
   \end{tabular}
   \par}
  \vskip 22pt
 }
 \par\noindent
 {\reset@font\small \par}
 \vskip 22pt
}
\makeatother

\title[Hybrid baryonic property insertion]{Hybrid analytic and machine-learned baryonic property insertion into galactic dark matter haloes}

\author[Moews, B. et al.]{
Ben Moews$^{1}$\thanks{E-mail: \href{mailto:bmoews@roe.ac.uk}{bmoews@roe.ac.uk}}
Romeel Dav\'e$^{1,2,3}$,
Sourav Mitra$^{4}$,
Sultan Hassan$^{5,2}$,
Weiguang Cui$^{1}$
\\
$^{1}$Institute for Astronomy, University of Edinburgh, Royal Observatory, Edinburgh EH9 3HJ, UK\\
$^{2}$Department of Physics and Astronomy, University of the Western Cape, Bellville, 7535, South Africa\\
$^{3}$South African Astronomical Observatories, Observatory, Cape Town 7925, South Africa\\
$^{4}$Surendranath College, 24/2 M. G. ROAD, Kolkata, West Bengal 700009, India\\
$^{5}$Center for Computational Astrophysics, Flatiron Institute, 162 5th Ave, New York, NY 10010, USA\\
}

\newcommand{\mufasa}{{\sc Mufasa}\xspace}
\newcommand{\simba}{{\sc Simba}\xspace}

\newcommand{\hmpc}{h^{-1}{\rm Mpc}}

\newcommand{\msolar}{\;{\rm M}_{\odot}}

\newcommand{\caesar}{{\sc Caesar}\xspace}

\newcommand{\mbh}{M_{\rm BH}}

\date{Accepted XXX. Received YYY; in original form ZZZ}

\pubyear{2020}

\usepackage{newtxtext,newtxmath}

\begin{document}
\label{firstpage}
\pagerange{\pageref{firstpage}--\pageref{lastpage}}
\maketitle

\begin{abstract}
While cosmological dark matter-only simulations relying solely on gravitational effects are comparably fast to compute, baryonic properties in simulated galaxies require complex hydrodynamic simulations that are computationally costly to run. We explore the merging of an extended version of the equilibrium model, an analytic formalism describing the evolution of the stellar, gas, and metal content of galaxies, into a machine learning framework. In doing so, we are able to recover more properties than the analytic formalism alone can provide, creating a high-speed hydrodynamic simulation emulator that populates galactic dark matter haloes in N-body simulations with baryonic properties. While there exists a trade-off between the reached accuracy and the speed advantage this approach offers, our results outperform an approach using only machine learning for a subset of baryonic properties. We demonstrate that this novel hybrid system enables the fast completion of dark matter-only information by mimicking the properties of a full hydrodynamic suite to a reasonable degree, and discuss the advantages and disadvantages of hybrid versus machine learning-only frameworks. In doing so, we offer an acceleration of commonly deployed simulations in cosmology.
\end{abstract}


\raggedbottom

\section{Introduction}
\label{sec:introduction}

Cosmological simulations are an invaluable part of theoretical research in cosmology, allowing for the implementation of new ideas and their testing against observations, as well as the detailed study of cosmological phenomena with large amounts of simulated data. The realistic modelling of processes when compared to observational data is a primary concern, as is the trade-off between simulation size and resolution~\citep{Dolag2008}.

While constraints on cosmological parameters are mostly directly tied to the overall matter distribution, observations can only directly probe the luminous baryonic component. Modelling the latter entails complex additional physical processes beyond gravity that result in much higher computational costs, which precludes parameter space explorations within $\sim$Gpc$^3$ volumes as needed for cosmological applications. It is, therefore, important to develop accurate frameworks to tie observable galaxy properties to the dark matter halo distribution.

Several approaches to solving this issue exist in the literature. One is based on abundance matching, in which the baryonic properties are tied to the stellar mass, which, in turn, is tied to the halo mass by assuming that rank ordering in mass is preserved. Here, satellites are extracted directly from the simulation, and rank ordering is implemented for all galaxies instead of modelling central and satellite galaxy relations separately. Conversely, halo occupancy distribution (HOD) modelling treats halo mass functions as an input and, assuming a satellite distribution, models the latter to match clustering constraints~\citep{Berlind2002}. The assumption of rank ordering, however, is not true in detail, and since there is no underlying physical model, it is not obvious that the often locally-calibrated relations apply at all the redshifts considered. That being said, with appropriate choices, it is possible to populate galaxies into dark matter haloes roughly in accord with observations.

Increasing in sophistication, semi-analytic models (SAMs) are another approach~\citep[see, e.g.,][]{White1991, Kauffmann1993, Cole1994}, which provides a full physical framework with increased computational cost, albeit still far cheaper than full hydrodynamic models. With appropriate parameter tuning, these can be calibrated to local relationships. Such models do, however, typically have a large number of free parameters that are difficult to constrain simultaneously, and so either tune parameters by hand, or else constrain only a subset of parameters to observations via a Markov chain Monte Carlo (MCMC) approach. As such, it is difficult to formally constrain the uncertainties in the physical parameters, as required for precision cosmology. Recently, machine learning has been employed to directly learn the relationship between dark matter haloes and baryonic properties within hydrodynamic simulations, which then allows populating those galaxy properties into dark matter haloes~\citep[see, e.g.,][]{Kamdar2016a, Kamdar2016b, Agarwal2018, Moster2020}. While stellar mass is quite accurately predicted, other properties such as star formation rates and gas contents have substantially poorer accuracy, which limits its usefulness in the age of increasingly precise surveys.

In this work, we introduce a novel framework that aims to marry the benefits of all of the above approaches. It is based on the {\it equilibrium model} introduced by~\citep{Dave2012}, a simple galaxy evolution framework whose free parameters correspond to baryon cycling, meaning the flows of material in and out of galaxies, which modern hydrodynamic simulations indicate is the main modulator of galaxy growth. With only eight free parameters, \citet{Mitra2015} show that this model can be MCMC-constrained to reproduce key relationships between halo mass ($M_h$), stellar mass ($M_*$), star formation rate (SFR), and metallicity ($Z$) of the galaxy population across much of cosmic time. \citet{Mitra2017} further demonstrate that it can reproduce the scatter in the $M_*-$SFR relation from expected inflow fluctuations owing to halo merging. The equilibrium model is far simpler than SAMs and hence substantially faster, which, together with fewer free parameters, enables MCMC to be applied over all parameters simultaneously.
The equilibrium model thus provides a way to tie galaxies to haloes that is fast and physically well-motivated, while providing formal Bayesian posterior information for cosmological error propagation.

That being said, the equilibrium model only directly predicts $M_*$, SFR, and $Z$, whereas cosmological applications often require a wider suite of baryonic properties such as neutral hydrogen content for HI intensity mapping. While the directly-predicted star formation and metallicity histories straightforwardly yield galaxy luminosities, accessing a wider suite of galaxy properties requires employing machine learning on hydrodynamic simulations that directly predict such properties. Although it is possible to use machine learning directly on the dark matter and thus bypassing the equilibrium model, \citet{Agarwal2018} report that their approach to predicting HI properties results in a large scatter relative to the true values. Factors like feature selection can, due to their respective relevance, play a role in these predictions, and a pure machine learning approach is not ruled out by this exploratory study.

When, however, additionally provided with true SFR and metallicity information, their machine learning model performs much better, even qualitatively recovering the second-parameter correlation between stellar mass, metallicity, and SFR (the so-called fundamental metallicity relation) \citep{Agarwal2018}. Thus, we can presume that by first applying the equilibrium model to dark matter-only simulations to predict key baryonic properties, and then feeding that information into the machine learning along with dark matter properties, it becomes possible to substantially improve the accuracy, or ease the process of reaching improved accuracies, of inserting galaxies into dark matter haloes, which is is our present goal.

In this work, we merge the equilibrium model into a machine learning framework to predict galaxy stellar and gas evolution within a merger tree. We demonstrate the effectiveness of this approach using the recently completed \simba cosmological hydrodynamic simulation~\citep{Dave2019}. Along the way, we implement extensions of the equilibrium model to account for the free-fall time within haloes, and enable the model to process largest-progenitor merger trees instead of just initial halo masses. By fusing this extended equilibrium model into an extremely randomised tree ensemble, a machine learning technique previously identified to perform well on the problem of baryonic property prediction, we advance the fields of both analytic galaxy evolution models for cosmological applications and machine learning for specialised tasks in the investigation of galaxy evolution~\citep{Kamdar2016a, Kamdar2016b, Agarwal2018, Jo2019, Hearin2020, Moster2020, Fluke2020}.

As a side benefit that we leave for future exploration, the MCMC-constrained equilibrium model makes quantitative predictions for key baryon cycling parameters that can inform and constrain galaxy formation models. For many cosmological applications, our approach offers a way to considerably reduce the computing resources required to predict comprehensive baryonic property sets, while leveraging information from both state-of-the-art Gpc-scale dark matter-only simulations and the latest hydrodynamic simulations. Large dynamic ranges, large Reynolds numbers, and highly supersonic flows make the modelling of baryonic physics in cosmological simulation numerically demanding when compared to the collisionless dynamics of dark matter in N-body simulations. Such efforts are, however, necessary to investigate theories of galaxy formation and evolution, as well as alternative cosmological models and their impact on galaxy populations~\citep{Vogelsberger2020}.

Machine learning techniques employed in the context of optimisation can be useful when trying to constrain parameters of cosmological simulations themselves, both in terms of observational data and formalisms like the equilibrium model. For parameter optimisation, MCMC methods are commonly used in astronomy, but can face certain limits in terms of expensive likelihood calculations and high numbers of dimensions. In recent years, multiple approaches have been developed to overcome these challenges, like renewed interest in cosmological applications of Approximate Bayesian Computation by \citep{Ishida2015} and~\citet{Akeret2015}, the Gaussbock algorithm by~\citet{Moews2020}, and Nested Sampling as introduced by~\citet{Skilling2006} and implemented into various packages~\citet{Liddle2006, Hobson2008, Feroz2009, Handley2015}.

The remainder of this paper is organised as follows. In Section~\ref{sec:background}, we provide the background on different types of cosmological simulations, the equilibrium model of galaxy evolution, an explanation of the type of machine learning model used in this work, and related recent research on machine learning for baryonic galaxy property prediction. Our methodology and data are described in Section~\ref{sec:methodology}, covering our newly proposed and subsequently implemented extensions of the equilibrium model, the creation of a hybrid prediction framework based on both the equilibrium model and machine learning, and the simulation from which we draw our dataset. We present our experimental setup and results, both for preliminary experiments with partial enhancements and a full experimental suite, in Section~\ref{sec:results}. Lastly, we discuss the results and implications of our work in Section~\ref{sec:discussion} and offer conclusions in Section~\ref{sec:conclusion}.

\section{Background}
\label{sec:background}

In this section, we provide the necessary background upon which this work builds, covering previous research as well as both the analytic and machine-learned part of our hybrid approach to inserting baryonic properties into N-body simulations. We provide an overview of cosmological simulations to investigate large-scale structure formation, namely N-body simulations, hydrodynamic simulations, and semi-analytic models,  as well as well-known simulations of each type, in Section~\ref{sec:simulations}. Following that, the equilibrium model, an analytic formalism of galaxy evolution, is described in Section~\ref{sec:equilibrium_model}, while Section~\ref{sec:trees} offers an overview of decision tree learning and ensembles, and Section~\ref{sec:related_work} describes related research on machine learning for baryonic property prediction.

\subsection{Structure formation and simulations}
\label{sec:simulations}

Generally, cosmological simulations can be split into three distinct approaches, which will be outlined in the following parts. As the simplest method for simulating the Universe, N-body simulations, which are also known as `pure-gravity' or `dark matter-only' simulations, employ dynamical systems of particles to calculate gravitational forces acting on them. This can be done either directly via numerical integration, or with the inclusion of general-relativistic effects to establish a scale factor $\alpha$ necessary for modelling the expansion of the Universe~\citep{Efstathiou1985}. Simulations of this kind played an essential role in establishing the $\Lambda$CDM model as the `standard model' of cosmology. Influential N-body simulations include the Millennium Simulation by~\citet{Springel2005a} and its Millennium-II successor as described by~\citet{Boylan_kolchin2009}, the Bolshoi simulation by~\citet{Klypin2011}, the subsequent MultiDark simulation by~\citet{Riebe2013}, the MICE Grand Challenge Lightcone Simulation as described by~\citet{Fosalba2015a, Fosalba2015b} and~\citet{Crocce2015}, and the EUCLID Flagship Simulation using PKDGRAV3~\citep{Hopkins2014, Potter2017}.

Since the baryonic content of the Universe can be described by treating gas in a continuous manner as an ideal fluid, hydrodynamic cosmological simulations using particle-based methods with discrete masses and grid-based methods with discrete spaces have been developed~\citep{Dolag2008}. Influential hydrodynamic simulations include MareNostrum Universe as described by~\citet{Hoeft2008}, the Illustris Simulation and its successor IllustrisTNG introduced by~\citet{Genel2014} and~\citet{Pillepich2018}, respectively, MassiveBlack-II by~\citet{Khandai2015}, EAGLE by~\citet{Schaye2015}, BlueTides as described by~\citet{Feng2016}, and Horizon-AGN by~\citet{Dubois2016}, as well as \mufasa and, more recently, its successor simulation \simba, with the latter being used in this work~\citep{Dave2016, Dave2019}.

Lastly, SAMs are built on models of baryonic physics to relate the hierarchical growth of dark matter haloes to galaxy population properties, as summarised by~\citet{Mitchell2018}, with an introduction to the field provided by~\citet{Baugh2006}. As SAMs combine (often simplified) physically motivated prescriptions with estimates of dark matter halo distributions and merger trees to calculate physical galaxy properties, and since the calculation of baryonic components with hydrodynamic simulations is computationally costly, large-volume investigations benefit from the SAM approach. An interface to N-body simulations exists through the use of dark matter halo merger trees extracted from such simulations as SAM inputs, as direct simulations are better suited for capturing non-linear structure formation than analytic methods like the Press-Schechter formalism~\citep{Press1974, Knebe2015}.

Influential models, while not an exhaustive list, include work by \citet{Somerville1999} and \citet{Somerville2008}, GALFORM as described by~\citet{Cole2000} and later recalibrated by~\citet{Baugh2018}, research by~\citet{Monaco2004} and~\citet{Kang2005}, GalICS and GalICS 2.0 by~\citet{Hatton2003} and~\citet{Cattaneo2017}, respectively, the Munich galaxy formation model by~\citet{Henriques2015}, the GAEA model by~\citet{Hirschmann2016}, and SAGE by~\citet{Croton2016}, as well as a broad review of galaxy formation theory by~\citet{Benson2010}. Notably, \citet{Neistein2012} present a method for turning hydrodynamic simulations into SAMs by transforming efficiencies in physical processes of galaxies into functions of $z$ and $M_{\text{halo}}$. While larger deviations for instantaneous properties like star formation rates are reported, this success increases the feasibility of machine learning methods that are trained on hydrodynamic simulations.

\subsection{The equilibrium model of galaxy evolution}
\label{sec:equilibrium_model}

Galaxies in hydrodynamic simulations have been observed to fluctuate around a self-regulatory relation on short timescales~\citep{Dutton2010}. In this context, the equilibrium model of galaxy evolution is an analytic formalism inspired by such simulations and based on the premise that galaxies are situated in slowly-evolving equilibria between inflow and outflow through accretion and feedback, as well as star formation, aiming to capture the evolution of galaxies in simulations \citep{Dave2012}. The equilibrium condition in the vicinity of which star-forming galaxies are seen to fall in such simulations is, with mass inflow rate $\dot{M}_{\text{in}}$, mass outflow rate $\dot{M}_{\text{out}}$ and star formation rate (SFR) $\dot{M}_*$,
\begin{eqnarray}
\dot{M}_{\text{in}} = \dot{M}_{\text{out}} + \dot{M}_*.
\label{eq:equilibrium_condition}
\end{eqnarray}
The reason for omitting a term for gas reservoirs, meaning the prevalence of molecular gas that is related to the star formation rate, is the finding by~\citet{Finlator2008} that the rate of change for such reservoirs is negligible in relation to the other terms in Equation~\ref{eq:equilibrium_condition}. While this constitutes a simplification, and despite changes in the gas reservoir having effects over short time frames, omitting the term still results in realistic galaxy growth when averaged over cosmological time frames. The interplay of mass inflow, outflow, and SFR in Equation~\ref{eq:equilibrium_condition} also bears resemblance to the reservoir model and the bath tub model~\citep{Bouche2010, Krumholz2012}. The notable difference is that the SFR is expressed as $(1-R)\dot{M}_*$ in these models, with $R$ as the (constant) gas recycling factor; in the equilibrium model, a time-dependent fitting formula is used for $R$ instead~\citep{Leitner2011}. The gas regulator model by~\citet{Lilly2013}, in the simplest form of which the specific star formation rate is set to the galaxy's specific accretion rate, is more accurate on shorter timescales, but increases the complexity of equations in return.

It also resembles SAMs in that both are analytic, but differs in the omission of merger trees (until now) and angular momentum conservation to cool gas. Using a Bayesian MCMC approach, \citet{Mitra2015} show that the model is well-suited to describe observations of scaling relations in galaxies from $z = 0$ to $z = 2$, with more recent investigations of $z = 0.5$ to $z = 3$ and including gas and dust observations~\citep{Saintonge2013, Mitra2017}. The equilibrium model rests on three central equations that describe the behaviour of the stellar, gas, and metal content over cosmological time frames. The SFR takes the form
\begin{eqnarray}
\dot{M_*} = \frac{\zeta \dot{M}_{\text{grav}} + \dot{M}_{\text{recyc}}}{(1 + \eta)},
\label{eq:SFR}
\end{eqnarray}
with $\eta \equiv \dot{M}_{\text{out}} / \dot{M}_*$ as the mass loading factor acting as an ejective feedback parameter, and $\zeta$ as the preventive feedback parameter describing the rate of growth of halo gas as the amount of halo-entering gas that does not reach the ISM, which can be defined by rearranging equation (\ref{eq:SFR}). $\dot{M}_{\text{recyc}}$ is parameterised as a function of $t_{\text{recyc}}$, which is the time frame it takes for ejected gas to be recycled. $\dot{M}_{\text{grav}}$ denotes the baryonic inflow into the dark matter halo as the infall into the halo that is assumed to be metal-free and derived from dark matter simulations, and $\dot{M}_{\text{recyc}}$ is the wind recycling parameter, meaning the re-accretion of previously ejected gas. The above equation is central to the equilibrium model's description of accretion and baryon cycling feedback steering the SFR, and can be derived as a reformulation of equation~\ref{eq:equilibrium_condition}. For a full derivation starting with the equilibrium condition, see~\citet{Dave2012}.  The metallicity in the ambient interstellar medium (ISM) gas $Z_{\text{ISM}}$ in the equilibrium model is, with $y$ as the  survey-derived metal yield,
\begin{eqnarray}
Z_{\text{ISM}} = y \frac{\text{SFR}}{\zeta \dot{M}_{\text{grav}}}.
\label{eq:Z_ISM}
\end{eqnarray}
The first part can be understood in the context of the metal enrichment rate, which is the yield times the SFR. Lastly, with the depletion time scaling as $t_{\text{dep}} \propto t_H M^{-0.3}_*$ for Hubble time $t_H$, and with the specific star formation rate $\text{sSFR}$, the dependence of a galaxy's gas fraction $f_{\text{gas}}$ on both $t_{\text{dep}}$ and $\text{sSFR}$ is described as
\begin{eqnarray}
f_{\text{gas}} = \frac{M_{\text{gas}}}{M_{\text{gas}} + M_*} = \frac{1}{1 + (t_{\text{dep}} \  \text{sSFR})^{-1}}.
\end{eqnarray}
Specifically, $t_{\text{dep}}$ represents the time frame in which gas in the ISM is converted into stars. Related work includes the modelling of the regulation of galactic star formation rates in disk galaxies by~\citet{Ostriker2010}, research on HI content of galaxies in hydrodynamic simulations by~\citet{Dave2013}, and the investigation of galactic outflows in cosmological zoom simulations~\citep{Angles-Alcazar2014}. For a more in-depth introduction, the reader is referred to the original paper by~\citet{Dave2012} or, for a broader overview of physical models of galaxy formation, \citet{Somerville2015}.

\subsection{Extremely randomised tree ensembles}
\label{sec:trees}

\begin{figure}
\includegraphics[width=\columnwidth]{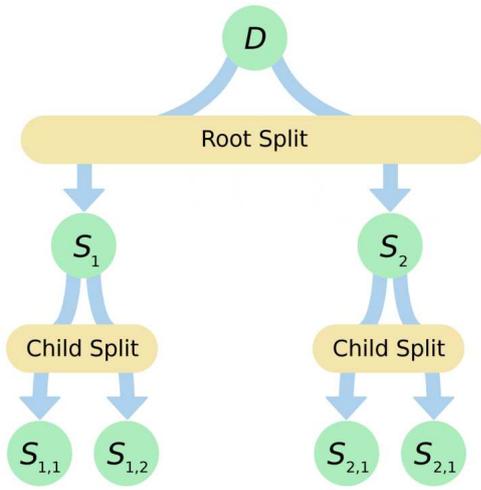}
\caption{Splitting process in extremely randomised trees. For a dataset $D$, which acts as the `root', the tree is built by generating binary splits to produce a deterministic flowchart, further splitting along the subsequent `child' nodes representing subsets until terminating in the end nodes as `leaves'.}
\label{fig:tree}
\end{figure}

Due to their simplicity and interpretability in terms of their decision-making process, \textit{decision trees} remain one of the most widely used machine learning algorithms~\citep[for a review, see][]{Wu2008}. As the name suggests, a decision tree is a hierarchical structure, starting with the input at the `root' and being subsequently split at nodes, usually in a binary manner, with final nodes corresponding to predicted values or classes being known as `leaves'. As such, these models can be viewed as generative models to create induction rules, making them an example of `white-box' models with clearly interpretable decision paths, the counterpart to black-box models like many types of neural networks. Figure~\ref{fig:tree} shows the splitting process in building said hierarchical structure. Notably, the use of multiple decision trees makes the resulting `forest' emerges as a black-box model, providing a trade-off of often better performance against a decrease in transparency~\citep{Guidotti2018}.

Introduced by~\citet{Breiman1984}, \texttt{CART} (short for `Classification And Regression Trees') is an early decision tree learning algorithm suitable for regression problems. \texttt{CART} traditionally makes use of the Gini impurity corresponding to Tsallis entropy, a generalised version of Boltzmann-Gibbs entropy in statistical mechanics,
\begin{eqnarray}
\mathrm{I}_G(p) = \sum_{i=1}^{N} p_i \sum_{k \neq i} p_k = \sum_{i=1}^{N} p_i (1 - p_i) = 1 - \sum_{i=1}^{N} p_i^2,
\end{eqnarray}
summing the probability of a given label, $p_i$, multiplied with the probably of a labelling error. Conversely, the \texttt{ID3} algorithm by~\citet{Quinlan1986} and its successor generally use the information gain,
\begin{eqnarray}
\mathrm{IG}(T, a) = - \sum_{i=1}^{N} \ p_i \ \mathrm{log}_2 \ p_i - \sum_{i=1}^{N} - \mathrm{p}(i | a) \ \mathrm{log}_2 \ \mathrm{p}(i | a),
\end{eqnarray}
for $N$ classes, a given label $i$, and $p_i$ as the percentages of each of these labels present in a splitting node. The expected information gain is then the mutual information, or reduction in entropy given by an optimal split. Given the iterative locally optimal splitting of datasets, building decision trees is a type of greedy algorithm~\citep{Quinlan1986}. For regression trees, the commonly used criterion is the mean squared error (MSE), which can be calculated, for a node $n$, associated data $D_n$, and samples $M_n$, as
\begin{eqnarray}
\mathrm{MSE}(D_n) = \frac{1}{M_n} \ \sum_{y \in D_n} (y - \bar{y}_n)^2, \ \mathrm{with} \ \bar{y}_n = \frac{1}{M_n} \sum_{y \in D_n} y.
\end{eqnarray}
Another common splitting criterion in regression is the mean absolute error (MAE), which replaces the squaring above with the modulus to obtain the non-squared difference. Using the MAE does, however, have the disadvantage of not being very punishing towards gross mispredictions with a linearly scaling error, making it suitable for datasets in which outliers can be safely disregarded. In machine learning, an \textit{ensemble} refers to a finite set of models, the output of which is used to produce the final output, for example by averaging or weighting the individual outputs. This is primarily done to combine a multitude of `weak learners' into a stronger predictive model, and to realise better generalisation. Two of the primary methods are \textit{boosting} and \textit{bootstrap aggregation} (also known as \textit{bagging}). The former incrementally refines an ensemble by sequentially training models on data points previously determined to be `hard', while the latter involves random draws with replacement from the dataset to create artificial training sets for the separate trees in an ensemble, averaging their output for predictions~\citep{Bishop2006}.

In the case of decision trees, the earliest and most wide-spread type of ensemble is the \textit{random forest}, an ensemble learning method first introduced by~\citet{Ho1995} and further expanded by~\citep{Breiman2001}. Making use of the random subspace method, trees are trained on bagging-style random samples of data points with replacement, combined with randomly sampled subsets of features, to increase accuracy and prevent overfitting to the training dataset~\citep{Ho2002}. For multivariate regression problems, such as the one tackled in this work, while one could build separate models for each output, training time concerns and correlations between output values for a given input usually lead to implementations using decision trees that predict all outputs~\citep{Dumont2009, Segal2011}.

Feature importances are commonly used to calculate the contribution of given inputs variables to the growth of tree-type machine learning models. These importances do, however, relate to which features are used most heavily in the construction of trees and do not necessarily connect with relationships in the underlying data, in this case physical importance. Correlated features can further bias the results, which is why these approaches should be treated with caution when it comes to their interpretation. For this reason, the development of statistical methods suitable to explore the underlying importances of interest is an active area of research, and we propose the use of ongoing developments in this field in future research~\citep{Strobl2007, Altmann2010, Fisher2019, Zhou2020}.

\begin{figure}
\includegraphics[width=\columnwidth]{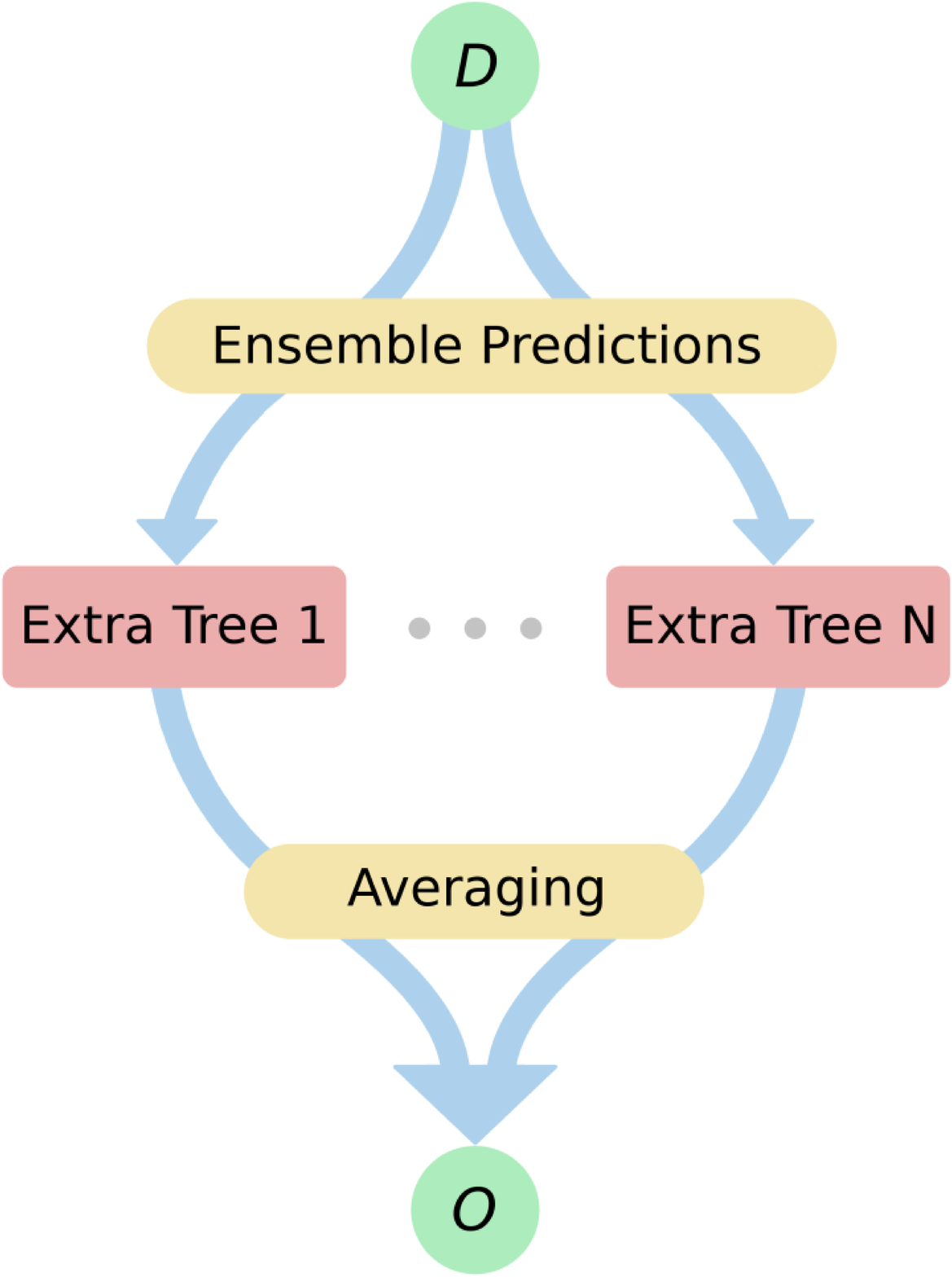}
\caption{Layout of predictions by extremely randomised trees as a previously trained ensemble of $N$ decision trees. For a given input dataset $D$ to produce predictions, the data is fed, in full, into each separate tree to generate predictions, which are then averaged to produce a final output $O$.}
\label{fig:ensemble}
\end{figure}

Adding an additional layer of randomness, \textit{extremely randomised trees} (usually shortened to \textit{extra trees}), while retaining the random subspace method targeting a random subset of features, discards the bootstrap sample, training each tree on the complete training dataset and choosing node splits based on a randomised selection instead of computing information-theoretically optimal splits~\citep{Geurts2006}. The flow of information in building an ensemble of extra trees can be seen in Figure~\ref{fig:ensemble}.

While complete randomness can be used to generate the splitting choices, split points for a given feature are usually randomly sampled from a uniform distribution of the feature's value range in the training dataset, followed by using the optimal choice among those randomly generated split proposals. For the splitting criterion, the mean squared error is used here, which translates to variance reduction for the feature selection. Algorithm~\ref{alg:extra_trees} describes this splitting process in extra trees as commented pseudocode.

\begin{algorithm}
\begin{footnotesize}
    \label{alg:extra_trees}
    \caption{Splitting in extremely randomised trees.}
    
    \KwData{\hspace{6pt}$S :=$ Local learning subset for node-splitting,\\
	\hspace{33pt}$k :=$ Number of attributes selected at each node,\\
	\hspace{33pt}$n :=$ Minimum sample size for splitting a node}
	\KwResult{$\hat{t} :=$ Optimal splitting choice for the given node}
	\textit{Check whether the subset $S$ fulfils splitting criteria}\\
	\uIf{$|S| < n$ \textnormal{\textbf{or}} $\forall c \in S : c$ const. \textnormal{\textbf{or}} $\mathrm{Tree}(S)$ const.}{
	    \textit{If not, return no splitting choice}\\
	    \Return None
	}
	\Else{
	\textit{Select $k$ non-constant attributes in $S$}\\
	    $\{a_1, \dots, a_k\} \sim \forall c \in S : c$ const.\\
	    \textit{Create an empty set to store random splits}\\
	    $T \longleftarrow \{ \}$\\
	    \textit{Create and store $k$ random splits for attributes}\\
	    \For{$i \in \mathbb{N}_{\leq k}$}{
	        $T \longleftarrow T \cup \{a_{\mathrm{cut}} \sim \mathcal{U}(\min(S_{a_i}), \max(S_{a_i})) \}$
	    }
	    \textit{Calculate and return the optimal splitting choice}\\
	    \Return $\hat{t} \in T$ s.t. $\mathrm{score}(t, S) = \max\limits_{i \in \mathbb{N}_{\leq k}}(\mathrm{score}(T_i, S))$
	}
	\vspace{10pt}
\end{footnotesize}
\end{algorithm}

Previous related research has explored the viability of different types of models for baryonic property prediction based on N-body simulations, allowing us to draw on existing research to determine the model of choice, as described below in Section~\ref{sec:related_work}~\citep{Kamdar2016b, Agarwal2018, Jo2019}.

\subsection{Machine learning and baryonic properties}
\label{sec:related_work}

While machine learning to paint dark matter haloes with galactic properties is sparse in the literature, several works have explored this topic so far. \citet{Kamdar2016a} introduced the application of machine learning techniques to semi-analytic cosmological simulations, training various algorithms to predict the total stellar mass $M_*$, the stellar mass in the bulge approximated via $M_{*, \mathrm{half}}$, and the central black hole mass $M_{\mathrm{BH}}$, as well as hot and cold gas masses, for each dark matter halo in the Millennium simulation at $z = 0$~\citep{Springel2005a}. Their research targets the prediction of these baryonic constituents based solely on dark matter halo merger trees as the training input, using the GADGET-2 algorithm described by~\citet{Springel2005b}, as well as the Tree-PM method by~\citet{Xu1995} to simulate gravitational interactions. In doing so, they extract both partial dark matter halo merger trees, with only the largest-mass progenitors, and the baryonic components. While hot gas masses, black hole bulge masses and stellar masses, both total and within the bulge, are predicted well with a slight overprediction of the bulge mass for lower-mass haloes, their approach suffers from poor predictive accuracy for cold gas masses.

In a follow-up project, \citet{Kamdar2016b} extend their previous machine learning framework for hydrodynamic simulations by using the public data release of the Illustris Simulation~\citep{Vogelsberger2014, Genel2014, Nelson2015}. Due to the previous success with extremely randomised trees in~\citet{Kamdar2016a}, which investigates decision trees, random forests, and the $k$-nearest neighbours algorithms for this problem, the same technique is employed again to predict $M_*$, $M_{\mathrm{BH}}$, gas mass $M_g$, SFR, and $g - r$ colour based on an input of dark matter halo properties without merger trees and a cosmology consistent with WMAP9~\citep{Hinshaw2013}. While recovering a similar population of galaxies via the used algorithm, a noticeable underprediction of the scatter is observed, with the possible explanation that the dark matter properties used as inputs do not contain enough information to learn the underlying physical processes. Notably, \citet{Kamdar2016b} make use of considerably more information in their inputs, namely the total dark matter subhalo mass, velocity dispersion, maximum circular velocity in the subhalo, the number of dark matter particles bound to the subhalo, and the three different spin components.

Similarly, \citet{Agarwal2018} perform experiments with decision trees, gradient-boosted trees, random forests, feed-forward neural networks, support vector regressors, the $k$-nearest neighbours algorithm, and extra trees, again finding that extra trees perform best for the prediction of baryonic properties when testing models on the hydrodynamic \mufasa simulation~\citep{Dave2016}. \citet{Agarwal2018} follow the aforementioned research by populating dark matter-only simulations with baryonic galaxy parameters via predicting $M_*$, SFR, metallicity $Z$, and both neutral ($M_{\mathrm{HI}}$) and molecular ($M_{\mathrm{H2}}$) hydrogen masses based on dark matter halo properties. The latter are, in this case, the dark matter halo mass, velocity dispersion, spin parameter, and nearby halo mass densities within radii at 200, 500, and 1000 kpc.

In applying this approach to the hydrodynamic \mufasa simulation introduced by \citet{Dave2016}, they observe the same underprediction of scatter around the mean relations as~\citet{Kamdar2016b} for the Illustris Simulation, and report that ensemble methods do not improve this result because none of the employed machine learning techniques reproduce the necessary scatter. They test a `meta-ensemble' by averaging the outputs of various machine learning algorithms, with methods like random forests and extremely randomised trees already being, as a combination of regression trees and bootstrap aggregation, an ensemble. Stacking or boosting would be more suitable for such an attempt to leverage the strength of different algorithms. \citet{Agarwal2018} also find that adding key baryonic parameters to the inputs, for example the SFR to predict $M_{\mathrm{HI}}$, improves the obtained results, which will become an important motivation for our present work in Section~\ref{sec:framework}.

Similarly, \citet{Moster2020} include the halo mass and peak halo mass, growth rate for both halo mass and peak halo mass, and the scale factor for halo mass, peak halo mass, and half-peak mass, as well as the virial radius, concentration parameter and spin parameter in their inputs to predict the stellar mass and SFR with a deep neural network using reinforcement learning. \citet{Jo2019} make use of the MultiDark-Planck~\citep[see][]{Klypin2016} and IllustrisTNG~\citep[see][]{Pillepich2018} simulations to estimate baryonic galaxy properties based on dark matter haloes and showing that results are largely compatible with SAMs. Extra trees are  chosen by~\citet{Jo2019}, too, further strengthening the evidence for this specific type of ensemble model in the literature. For this reason, given the previous determination of the model most suitable for the problem at hand, we focus on the use of extra trees in this work.

In related research, \citet{Lucie-Smith2018} show that random forests can be used to classify whether dark matter particles will be part of haloes in a given mass range, matching the predictions of common spherical collapse approximations. Interestingly, the application of modern machine learning to cosmological simulations is still relatively sparse in the literature, although the number of contributions is growing, with close demonstrations with regard to large-scale structures tackling the prediction of cosmological parameters with 3D simulations of the cosmic web by~\citet{Ravanbaksh2016}, as well as the creation of cosmic web simulations and synthetic galaxies~\citep{Ravanbaksh2017, Rodriguez2018, Fussell2019}.

\section{Methodology and data}
\label{sec:methodology}

In this section, we provide details on the methodological considerations and data sources contributing to this work. We propose extensions to the equilibrium model of galaxy evolution, covering the inclusion of free-fall time and merger trees, and describe them in Section~\ref{sec:extension}. Following this, Section~\ref{sec:framework} outlines the hybrid prediction approach we create from merging our extended equilibrium model into a machine learning framework. In Section~\ref{sec:data}, we describe the cosmological simulation we make use of in this work, as well as the extraction of the data used in the presented experiments.

\subsection{Extension of the equilibrium model}
\label{sec:extension}

We include some minor improvements to the equilibrium model presented in \citet{Dave2012} and \citet{Mitra2015}.  First, we introduce a time delay between the accretion onto the halo and the accretion onto the ISM, given by the free-fall time of the halo at the given redshift,
\begin{eqnarray}
t_{\mathrm{ff}} = \sqrt{\frac{3 \pi}{32 G \rho}} \ , \ \mathrm{where} \ \rho = 200 \rho_{\mathrm{crit},z=0} (1 + z)^3.
\end{eqnarray}

More significantly, another novel addition to the equilibrium model is that the halo growth rate is now computed based on largest-progenitor merger trees. The original equilibrium model of \citet{Dave2012} employed the fitting formula from~\citet{Dekel2009} for the average growth rate of $M_h$ as
\begin{eqnarray}
\dot{M}_h = 6.6 \left( \frac{M_h}{10^{12}} \right)^{1.15} (1 + z)^{2.25} f_{0.165} M_\odot \mathrm{yr}^{-1},
\end{eqnarray} 
which was based on the assumption of a flat Universe with a fluctuation normalisation parameter of $\sigma_8 = 0.8$, mass-dominated by cold dark matter and with 72\% dark energy content. \citet{Mitra2017} extended this to include inflow fluctuations based on a parameterisation from the Millenium simulation.

In our new version, we employ halo merger trees to compute the halo growth rate. The dark matter particle mass resolution is $\approx 10^8 M_\odot$, which is not ideal for probing the very earliest phases of galaxy growth, but for this initial proof of concept it suffices. For each halo, between each of the 114 time steps, we compute the average growth rate during that step.  If the growth rate is negative, we employ `backwards capping' and wait until the halo increases in mass at a later step, and take the average growth rate over all steps until it becomes positive.

Owing to the finite number of merger tree outputs resulting in up to hundreds of Myr between time steps, we augment this growth using the same formalism as \citet{Mitra2017} to account for short-timescale inflow fluctuations, with the limit that the fluctuations cannot grow the halo more than the amount for the entire time step.  In this way, we account for both the individual long-term growth history of haloes, as well as (statistically) the fluctuations that may drive the scatter in galaxy scaling relations. This provides a more realistic description of haloes taken directly from an N-body simulations, which we later explore in Section~\ref{sec:full_experiments}. 

\subsection{Creating a hybrid prediction framework}
\label{sec:framework}

\begin{figure}
\includegraphics[width=\columnwidth]{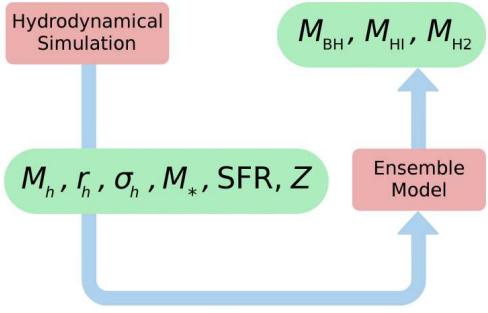}
\caption{Training process of the machine learning component of the hybrid framework presented here. The depicted workflow shows the training of an ensemble model based on the dark matter halo mass ($M_h$), dark matter half-mass radius ($r_h$), dark matter halo velocity dispersion ($\sigma_h$), stellar mass ($M_*$), star formation rate (SFR), and metallicity ($Z$) of a galaxy within a hydrodynamic simulation to predict the corresponding black hole mass ($M_{\mathrm{BH}}$), neutral hydrogen ($M_{\mathrm{HI}}$) mass, and molecular hydrogen mass ($M_{\mathrm{H2}}$).}
\label{fig:training}
\end{figure}

In order to create a hybrid framework making use of both analytic formalisms and machine learning, we implement two modules, the first of which is an extra trees ensemble as introduced in Section~\ref{sec:trees}. Figure~\ref{fig:training} shows the training workflow of this model, starting with input values from a hydrodynamic simulation such as \simba~\citep{Dave2019}. The depicted setup is related to the previously introduced conjecture that the inclusion of additional baryonic properties improves the accuracy of predictions for remaining properties~\citep{Agarwal2018}. As N-body simulations, while relatively fast to compute, provide only dark matter properties, the equilibirum model offers a way to estimate a subset of baryonic properties as additional machine learning inputs.

In our case, the dark matter halo mass ($M_h$), dark matter half-mass radius ($r_h$), dark matter halo velocity dispersion ($\sigma_h$), stellar mass ($M_*$), star formation rate (SFR), and metallicity ($Z$) of a galaxy are used as inputs to predict the corresponding black hole mass ($M_{\mathrm{BH}}$), neutral hydrogen ($M_{\mathrm{HI}}$), and molecular hydrogen ($M_{\mathrm{H2}}$). Apart from the hybrid approach for additional baryonic inputs, this presents an additional difference to previous research, specifically to~\citep{Kamdar2016b} not predicting neutral and molecular hydrogen but the total gas mass. Due to working on the basis of a different hydrodynamic simulation, the latter also use additional dark matter inputs, namely the three spin components, the maximum circular velocity in the subhalo, and the number of dark matter particles bound to the subhalo, but omit $r_h$ as an input. 

Our set of inputs more closely resembles work by~\citet{Agarwal2018}, but the latter make use of the halo spin instead of $r_h$. As $r_h$ tightly connects with halo concentration, and thus the halo formation time which dominates the scatter in the $M_*-M_h$ relation, we expect that $r_h$ provides an orthogonal dimension to the other provided properties. Additionally, they include nearby halo mass densities within radii at 200, 500, and 1000 kpc, which translates to environmental data, and do not predict $M_{\mathrm{BH}}$. Notably, though, they use arrays of $M_h$ for both the present and the five immediately preceding snapshot in their inputs, thus including merger tree information directly into the machine learning model. This consideration is later adapted into the full merger tree experiments in Section~\ref{sec:full_experiments}.

\begin{figure}
\includegraphics[width=\columnwidth]{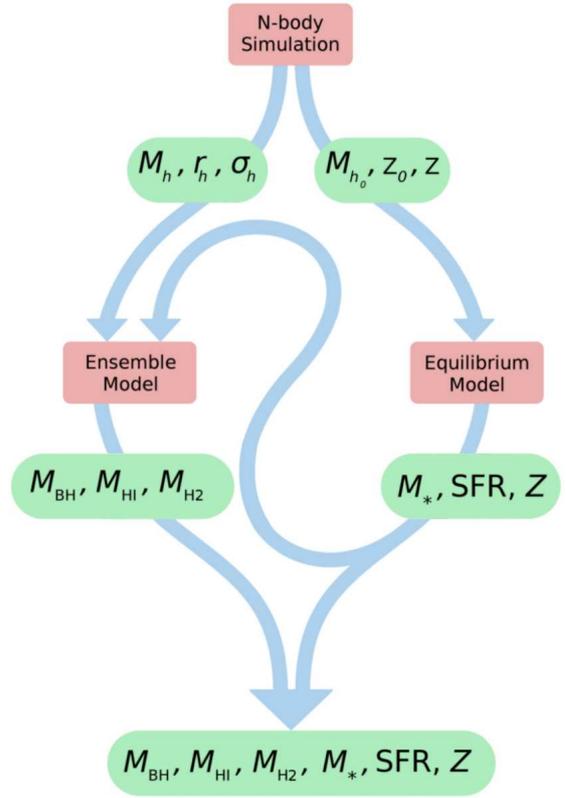}
\caption{Prediction with the full hybrid analytic and machine learning framework presented in this work. In the shown workflow, along the right path, the merger tree-based initial halo mass ($M_{h_0}$) of a given galaxy, as well as initial and final redshifts ($z_0, z$) for the same merger trees, are fed into our modified version of the equilibirum model to produce the corresponding stellar mass ($M_*$), star formation rate (SFR), and metallicity ($Z$). These values, along the left path and together with the dark matter halo mass ($M_h$), half-mass radius ($r_h$), and dark matter halo velocity dispersion ($\sigma_h$), are then used by the previously trained ensemble model to predict the black hole mass ($M_{\mathrm{BH}}$), neutral hydrogen ($M_{\mathrm{HI}}$), and molecular hydrogen ($M_{\mathrm{H2}}$), as well as updated outputs of the values predicted by the equilibrium model.}
\label{fig:prediction}
\end{figure}

Importantly, this means that our approach runs on a reduced amount of information when compared to previous research on methods using only machine learning, with only three basic dark matter inputs fed into the framework. This is partly counterbalanced by the inclusion of merger tree data in the equilibrium model introduced in Section~\ref{sec:equilibrium_model}, featuring the extensions described in Section~\ref{sec:extension}. As shown in Figure~\ref{fig:prediction}, the merger tree-based initial halo mass ($M_{h_0}$) of a galaxy, as well as initial and final redshifts ($z_0, z$) for a given merger tree, are fed into our modified version of the equilibirum model to produce the corresponding stellar mass ($M_*$), star formation rate (SFR), and metallicity ($Z$).

These outputs, together with the previously described dark matter inputs, are then used as inputs to the trained ensemble model, predicting the full set of baryonic properties of a given galaxy. The baryonic properties generated by the equilibrium model are used both as an input and as target variables of the extra trees ensemble to further refine the results based on the combined inputs. Step-wise, our model thus works in the following way:

\begin{itemize}
\item Train the machine learning model on \simba data
    \begin{itemize}
    \item Inputs: $M_*$, SFR, $Z$, $M_h$, $r_h$, $\sigma_h$
    \item Outputs: $M_*$, SFR, $Z$, $M_{\mathrm{BH}}$, $M_{\mathrm{HI}}$, $M_{\mathrm{H2}}$
    \end{itemize}
\item Derive baryonic outputs of the equilibrium model
    \begin{itemize}
    \item Inputs: $M_{h_0}$, $z_0$, $z$
    \item Outputs: $M_*$, SFR, $Z$
    \end{itemize}
\item Predict baryonic quantities with the trained model
\end{itemize}

In doing so, we make use of the equilibrium model to create the additional baryonic inputs necessary to run a machine learning model relying on them, while only requiring dark matter properties commonly found in comparatively fast-running N-body simulation as inputs for the completed framework.

\subsection{Data from \simba}
\label{sec:data}

The dataset is extracted from the m100n1024 run of the \simba simulation~\citep{Dave2019}, which has a volume of 100 $\hmpc$ with $1024^3$ dark matter particles and $1024^3$ gas elements. It assumes a Planck Collaboration XIII (2016) concordant cosmology of $\Omega_m = 0.3$, $\Omega_\Lambda = 0.7$, $\Omega_b = 0.048$, $H_0 = 68\ km\ s^{-1}\ Mpc^{-1}$, $\sigma_8 = 0.82$, and $n_s = 0.97$. We refer to \citet{Dave2019} for its detailed physical models for baryon processes. This simulation starts at $z = 249$, with $\sim$ 150 outputs spanning from $z = 30$ to zero. 

Haloes are identified on the fly by a 3D friends-of-friends algorithm within \textsc{GIZMO}, with a linking length set to 0.2 times the mean inter-particle spacing and without the consideration of subhaloes. We identify galaxies with a YT-based package, \caesar \footnote{\url{https://github.com/dnarayanan/caesar}}, which uses a 3D friends-of-friends galaxy finder that assumes a spatial linking length of 0.0056 times the mean inter-particle spacing (equivalent to twice the minimum softening length). Black holes that are gravitationally bound, as well as gas elements with a minimum SF threshold density of $n_H > 0.13 \text{ H atoms cm}^{-3}$, are attached to the host galaxy with the same linking length. We treat the most massive black hole in the galaxy as the central black hole, with its mass as $M_\mathrm{BH}$. The neutral and molecular hydrogen of the galaxy are calculated based on its gas properties; these are computed self-consistently in \simba assuming self-shielding from~\citet{Rahmati2013} and atomic/molecular separation based on the subgrid model of \citet{Krumholz2012}.

Galaxies and haloes are cross-matched in post-processing within \caesar, and the most massive galaxy close to the halo minimum potential centre is assigned as the central galaxy. The merger tree of a galaxy, instead of a halo, is built by tracking the unique star particle IDs, while the most massive progenitor is treated as the main progenitor of the descent, which we use for tracking galaxy evolution here. As there is a one-to-one connection between the central galaxy and its host halo, it is simple to have the host halo information attached to its merger tree. We constrict our data to final halo masses of $\mathrm{log}_{10} (M_h) \in [11, 14]$ and make sure logarithms of values do not lead to unsuitable infinities, but do not make use of any ways of further restricting the dataset that could lead to better fits, either for the data extracted from the \simba simulation or our framework's predictions, in order to present generalisable results. In doing so, we generate a dataset of $30,555$ haloes with corresponding merger trees and baryonic properties to create training and test sets from, as well as a separate dataset for the baryon cycling parameter optimisation via MCMC, as described in Section~\ref{sec:splining}.

\section{Experiments and results}

\label{sec:results}

In this section, we describe the experiments performed to evaluate the performance of our approach to baryonic property insertion, both in a half-way setup and a full-scale experiment the extensions proposed and described in this work. A preliminary experimental run is covered in Section~\ref{sec:splining}, without merger trees and dark matter halo mass variability, but including a free-fall time modification of the equilibrium mode and relying on splining outputs of the equilibrium model to build a relation between initial and final halo mass in lieu of using largest-progenitor merger trees. After including the use of merger trees in the equilibrium model, we present the final results in Section~\ref{sec:full_experiments}.

\subsection{Preliminary splining and free-fall time}
\label{sec:splining}

\begin{figure}
\includegraphics[width=\columnwidth]{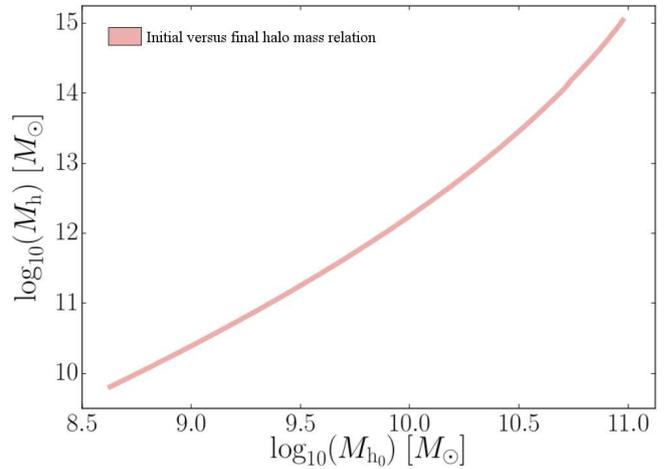}
\caption{Splining of the relation between initial halo mass ($M_{h_0}$) and final halo mass ($M_h$) of galaxies in the equilibrium model. Based on the halo masses extracted from the \simba simulation and a redshift of $z = 10$, 100 equidistantly spread initial halo mass values are fed into the model to cover the corresponding final halo mass range, with the result being splined to approximate a continuous look-up function.}
\label{fig:splining}
\end{figure}

In the first step of our experiments, we include the free-fall time, as described in Section~\ref{sec:extension}, into the equilibrium model, but omit merger trees for now. We represent the baryon cycling parameters by nine free variables to assess their behavior with halo masses and redshifts \citep[see also][]{Mitra2015,Mitra2017},
\begin{eqnarray}
\eta &=& \left(\frac{M_h}{10^{\eta_1+\eta_2(1+\mathrm{min}(z, 2))}}\right)^{\eta_3},\\
t_{\rm rec} &=& \tau_1 t^{\tau_2} \left(\frac{M_h}{10^{12}}\right)^{\tau_3},\\
\zeta_{\rm quench} &=& 
\begin{cases}
\left(\frac{M_h}{M_q}\right)^{\zeta_1 (1 + z)}, & \text{if}\ M_h > M_q \\
1, & \text{else}
\end{cases},
\end{eqnarray}
where $\eta$ is the ejective feedback parameter, $t_{\rm rec}$ is the wind recycling time at a given time $t$, and $\zeta_{\rm quench}$ is the quenching feedback parameter, with a corresponding quenching mass $M_q = 10^{12}\msolar (\zeta_2+\zeta_3 z)$. We note that the parameterisations have changed slightly from \citet{Mitra2015}, as these were found to give more reasonable extrapolations to higher redshifts.
As in \citet{Mitra2015}, these free parameters $\{\eta_1,\eta_2,\eta_3,\tau_1,\tau_2,\tau_3,\zeta_1,\zeta_2,\zeta_3\}$ are then constrained using a Bayesian MCMC analysis with $N = 500$ walkers against recent observations on the $M_*-M_{\mathrm{h}}$ relation by~\citet{Behroozi2019}, $M_*-Z$ relation~\citep[combined from][]{Andrews2013,Zahid2014,Ly2016,Sanders2018}, and $M_*-$SFR relation by~\citet{Speagle014} at redshifts $z \in {0, 1, 2}$. The best-fit values we obtain from the resulting analysis are listed in Table~\ref{tab:table_cycling}, and are subsequently used for the equilibrium model component of our framework. The match between our model predictions and observed datasets are quite similar to earlier results on this MCMC fitting approach by \citet{Mitra2015} and \citet{Mitra2017}, and we refer the reader to those papers for a more detailed description.

\begin{table}
\caption{Baryon cycling parameters for the equilibrium model used in the analytic formalism part of our framework. The best-fit values are achieved through a Bayesian MCMC estimation for ejective feedback parameters ($\eta$), wind recycling parameters ($\tau$) and quenching feedback parameters ($\zeta$).}
 \label{tab:table_cycling}
 \begin{tabular*}{\columnwidth}{@{}l@{\hspace*{5pt}}l@{\hspace*{5pt}}l@{\hspace*{5pt}}l@{\hspace*{5pt}}l@{\hspace*{5pt}}l@{\hspace*{5pt}}l@{\hspace*{5pt}}l@{\hspace*{5pt}}l@{}}
  \hline
  $\eta_1$ & $\eta_2$ & $\eta_3$ & $\tau_1$ & $\tau_2$ & $\tau_3$ & $\zeta_1$ & $\zeta_2$ & $\zeta_3$\\[2pt]
  \hline
  $10.822$ & $0.405$ & $-1.517$ & $3.184$ & $-2.161$ & $-1.381$ & $-0.229$ & $1.122$ & $0.007$\\[2pt]
  \hline
 \end{tabular*}
\end{table}

\begin{figure}
\includegraphics[width=\columnwidth]{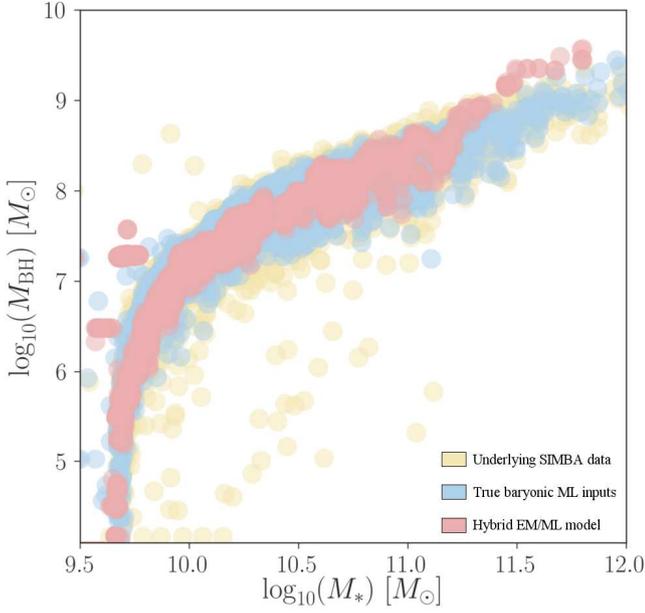}
\caption{Scatter plot for the restricted experimental run, with reduced scatter typical of machine learning approaches. The figure shows results for stellar mass ($M_*$) versus black hole mass ($M_{\mathrm{BH}}$), with true \simba data plotted in yellow, results of an extremely randomised tree ensemble with additional baryonic inputs shown in blue, and results for the preliminary test of the hybrid analytic and machine learning framework without some of the extensions introduced in this work shown in red.}
\label{fig:combined_predictions}
\end{figure}

We choose a redshift of $z = 10$ as a baseline value and feed initial halo masses $M_{h_0}$ into the equilibrium model using the \citet{Dekel2009} mean accretion rate (not including fluctuations) to generate final halo masses covering values on a closed interval $\mathrm{log}_{10} (M_h) \in [11, 14]$ for later evaluation against \simba data. Using the mean accretion rate allows a one-to-one mapping of initial and final halo masses that is on average correct, and enables us to identify the starting halo mass for any halo in the merger tree. Figure~\ref{fig:splining} shows the resulting spline as a continuous look-up function for initial versus final halo mass. For training and test sets, we split our dataset in a four-to-one ratio via random subsampling without replacement, resulting in a training set of $24,444$ and a test set of $6,111$ examples.

\begin{figure*}
\includegraphics[width=\textwidth]{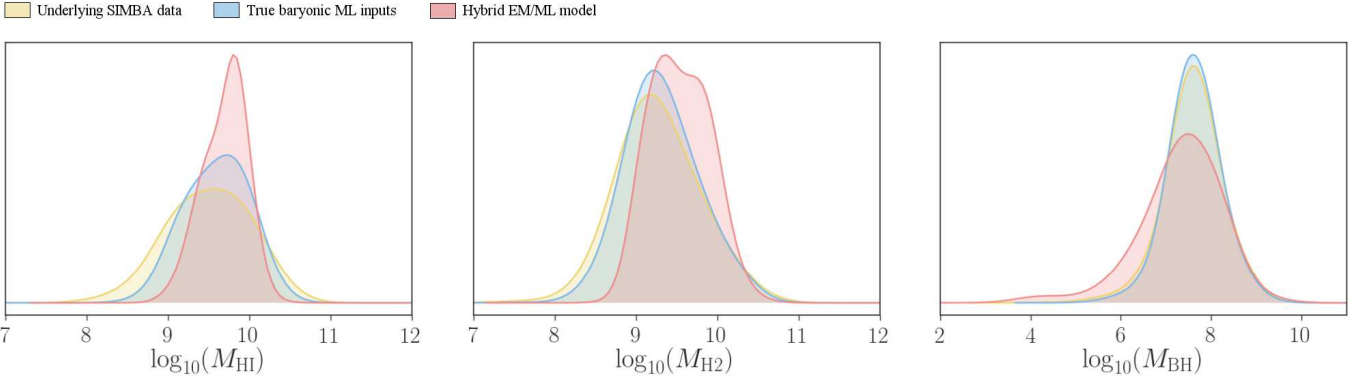}
\caption{Density plots for the restricted experimental run. The panel show results for neutral hydrogen ($M_{\mathrm{HI}}$), molecular hydrogen ($M_{\mathrm{H2}}$), and black hole mass ($M_{\mathrm{BH}}$). In all three panels, the true underlying \simba target data is plotted in yellow. Results of an extremely randomised tree ensemble with additional true baryonic inputs from \simba, mimicking a hypothetical `perfect' equilibrium model by receiving the actual target values for these inputs, are shown in blue and lead to a green tint when fitting the underlying data. Lastly, results for the preliminary test of the hybrid analytic and machine learning framework without some of the extensions introduced in this work are shown in red.}
\label{fig:histograms}
\end{figure*}

We train an extra trees ensemble as described in Section~\ref{sec:trees} and Section~\ref{sec:framework} and shown in Figure~\ref{fig:training}, feeding both dark matter properties and baryonic properties predicted via the equilibrium model into it. Specifically, this means that an array of \{$M_{h_0}$, $z_0$, $z$\} values are used by the equilibrium model to predict the corresponding array of \{$M_*$, SFR, $Z$\} values as shown in Figure~\ref{fig:prediction}, with $z_0 = 10$, $z = 0$, and $M_{h_0}$ predicted from the aforementioned splining of \simba halo masses. These output values, together with \simba values \{$M_h$, $r_h$, $\sigma_h$\}, are then used by the ensemble to predict \{$M_{\mathrm{BH}}$, $M_{\mathrm{HI}}$, $M_{\mathrm{H2}}\}$. In addition, we also train two additional extra trees ensembles for the purpose of comparison. First, we train one model that is allowed to use true underlying target values from \simba instead of equilibrium model outputs in order to gauge the effect the inclusion of these partial baryonic properties has on the quality of the results. Effectively, this machine learning-only setup emulates the assumption of perfect predictions by the equilibrium model. Secondly, we also train one model that disregards any baryonic input information, predicting solely based on \simba values \{$M_h$, $r_h$, $\sigma_h$\} as a pure machine learning baseline.

Due to the limited information and fixed redshift value for looking up initial halo masses from a spline, we can expect results to feature some irregularities. Figure~\ref{fig:combined_predictions}, depicting a two-dimensional plot for $M_*$ and $M_{\mathrm{BH}}$ of this restricted setup as an easy-to-eyeball combination with a noticeable bend, demonstrates this by showing a lack of scatter in $M_{\mathrm{BH}}$ for larger values of $M_*$. For an closer visual analysis of the results, taking a look at separate variables can be useful to determine the level of overconstrained or underconstrained distributions and eventual missed or superfluous multimodal features. Density plots of all output values are shown in Figure~\ref{fig:histograms}, and are further discussed in Section~\ref{sec:discussion}. The data shown in Figure~\ref{fig:combined_predictions} can be summarised as follows, and follows the same colour scheme for subsequent Figures~\ref{fig:histograms} and~\ref{fig:histograms2} further below:

\begin{itemize}
\item Yellow: True underlying \simba data, meaning the target values taken directly from the cosmological simulation instead of the output of a predictive model, as a comparison baseline
\item Blue: Machine learning-only results when, as an idealised scenario, feeding true underlying \simba data for \{$M_*$, SFR, $Z$\} into the extra trees ensemble instead of using equilibrium model estimates, thus mimicking a hypothetical `perfect' equilibrium model
\item Red: Hybrid model results, predicting \{$M_*$, SFR, $Z$\} via the equilibrium model and using the extra trees machine learning model to estimate the full set of baryonic target properties
\end{itemize}

While visual inspections provide a reasonable overview, stricter statistical validation is necessary to assess the results. For this purpose, we calculate the coefficient of determination,
\begin{eqnarray}
R^2(x, y) = 1 - \frac{\sum_{i = 1}^n (x_i - y_i)^2}{\sum_{i = 1}^n (x_i - \bar{x})^2},
\end{eqnarray}
with $x$ and $y$ as the true target values and their predictions generated by our framework, respectively, representing the proportion of the dependent variable's variance that can be explained through the independent variable. As such, it offers a way to quantify the performance of a model's replication of observed outcomes and, while having an upper limit of one, features no lower limits for models that perform arbitrarily badly. Specifically, negative values for non-linear functions, as is the case in our work, mean that the data's mean provides a better fit than the predictions in question. In addition, we also calculate Pearson's correlation coefficient,
\begin{eqnarray}
\rho(x, y) = \frac{\sum_{i = 1}^n (x_i - \bar{x}) (y_i - \bar{y})}{\sqrt{\sum_{i = 1}^n (x_i - \bar{x})^2 (y_i - \bar{y})^2 }},
\end{eqnarray}
as a measure that can be employed together with $R^2$ \citep[see, for example,][for related research]{Agarwal2018} to provide additional insight. It measures the linear correlation two variables, in our case the true target values and the framework's predictions, and is limited to $\rho(x, y) \in [-1, 1]$.

\begin{table}
\caption{Statistical validation for the restricted experimental run. The table lists the coefficient of determination ($R^2$) and Pearson's correlation coefficient ($\rho$) for different setups. The column denoted as `True' shows results for the prediction of neutral hydrogen ($M_{\mathrm{HI}}$), molecular hydrogen ($M_{\mathrm{H2}}$), and black hole mass ($M_{\mathrm{BH}}$) when feeding true underlying \simba target values for stellar mass ($M_*$), and star formation rate (SFR) into the machine learning model, while the column under `ML' shows results for excluding ${M_*, SFR, Z}$ from the inputs, predicting only based on dark matter halo information. The column under `Hybrid' shows the results when using the equilibrium model without merger tree information for these baryonic inputs, and predicting these as well, for an invariant initial redshift of $z = 0$ and initial halo masses predicted from splined equilibrium model results.}
 \label{tab:table_1}
 \begin{tabular*}{\columnwidth}{@{}l@{\hspace*{10pt}}l@{\hspace*{10pt}}l@{\hspace*{15pt}}l@{\hspace*{10pt}}l@{\hspace*{15pt}}l@{\hspace*{10pt}}l@{}}
  \hline
  & \multicolumn{2}{@{}l}{\textbf{True}} & \multicolumn{2}{@{}l}{\textbf{ML}} & \multicolumn{2}{@{}l}{\textbf{Hybrid}}\\[2pt]
  \hline
  Variable & $R^2$ & $\rho$ & $R^2$ & $\rho$ & $R^2$ & $\rho$\\[2pt]
  \hline
  $M_{{\mathrm{{HI}}}}$ & $0.5560$ & $0.7497$ & $0.4125$ & $0.6538$ & $0.2563$ & $0.5343$\\[2pt]
  $M_{{\mathrm{{H2}}}}$ & $0.7743$ & $0.8800$ & $0.7456$ & $0.8635$ & $0.5457$ & $0.7631$\\[2pt]
  $M_{{\mathrm{{BH}}}}$ & $0.7207$ & $0.8832$ & $0.6354$ & $0.8765$ & $0.5980$ & $0.8517$\\[2pt]
  $M_*$ & $-$ & $-$ & $-$ & $-$ & $0.7286$ & $0.9621$\\[2pt]
  $\mathrm{{SFR}}$ & $-$ & $-$ & $-$ & $-$ & $0.7266$ & $0.8663$\\[2pt]
  $Z$ & $-$ & $-$ & $-$ & $-$ & $-7.4543$ & $0.4847$\\[2pt]
  \hline
 \end{tabular*}
\end{table}

The results for these calculations are listed in Table~\ref{tab:table_1}, with the column under `True' corresponding to a machine learning model receiving true underlying $M_*$, SFR, and $Z$ values from \simba (coloured yellow in Figures~\ref{fig:combined_predictions} and~\ref{fig:histograms}), and the column under `ML' to predictions based on feeding only dark matter halo features into the machine learning model while disregarding the equilibrium model (coloured blue in Figures~\ref{fig:combined_predictions} and~\ref{fig:histograms}). The column under `Hybrid' refers to predictions using the equilibrium model for these baryonic inputs, albeit with a fixed redshift and initial halo masses $M_{h_0}$ produced by the continuous look-up function via a spline described in this section and shown in Figure~\ref{fig:splining} (coloured red in Figures~\ref{fig:combined_predictions} and~\ref{fig:histograms}).

\subsection{Inclusion of merger tree information}
\label{sec:full_experiments}

In the next step, we include the remaining extension of the equilibrium model to run a full test suite for our approach. This means that merger tree information described in Section~\ref{sec:extension} is incorporated into the hybrid model, both in terms of the internal use of merger trees by the equilibrium model and the five most recent halo masses as part of the inputs as in~\citet{Agarwal2018}. For this experiment, we make use of the same dataset as previously, with training and test set examples numbering $24,444$ and $6,111$, respectively, to enable an as-accurate-as-possible measurement of the impact that the inclusion of merger tree data, specifically the associated halo masses, have on the hybrid framework.

\begin{figure*}
\includegraphics[width=\textwidth]{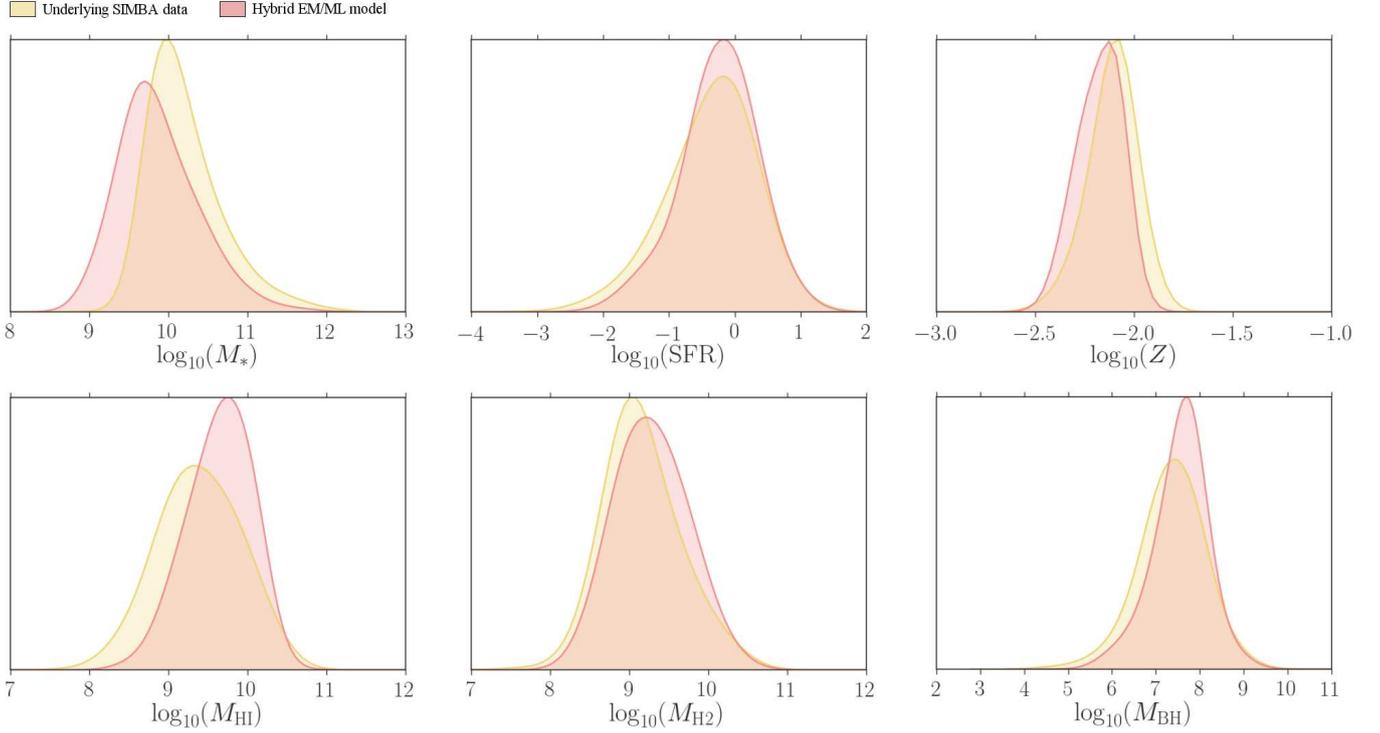}
\caption{Density plots for the full experimental run including merger trees. The panel show results for stellar mass ($M_*$), star formation rate (SFR), metallicity ($Z$), neutral hydrogen ($M_{\mathrm{HI}}$), molecular hydrogen ($M_{\mathrm{H2}}$), and black hole mass ($M_{\mathrm{BH}}$). In all six panels, the true underlying \simba data is plotted in yellow,} and results for the test of the hybrid analytic and machine learning framework with all extensions introduced in this work are shown in red.
\label{fig:histograms2}
\end{figure*}

As previously in Section~\ref{sec:splining}, we use an extra trees ensemble trained on the \simba-extracted data described in Section~\ref{sec:data}, as shown in Figure~\ref{fig:training}, and use the already MCMC-fitted baryon cycling parameters. Unlike in these preliminary experiments, however, we feed full largest-progenitor merger trees instead of just the respective initial halo mass estimates into the equilibrium model, allowing the model to steer more closely along each dark mater halo's true evolutionary history. As the prediction of $Z$ stays too constrained when making use of equilibrium model outputs, we instead predict the property directly from the other outputs. Fitting tailored parameters for the given larger dataset could possibly provide a more accurate fit, but this would incur a significantly higher computational cost due to the increased wall time of including merger trees. In addition, using pre-fitted parameters relying on a previous and smaller dataset also provides a use case for further applications, as future research is planned to apply our framework to N-body simulations that are unable to provide the target values for these fitting procedures.

Density plots of all output values are shown in Figure~\ref{fig:histograms2}, and are further discussed in Section~\ref{sec:discussion}. Again, we compare a hypothetical `perfect' equilibrium model by using true underlying \simba values instead of mode-outputted values for $M_*$, SFR, and $Z$, and find that the respective statistical key performance indices remain at a very similar level to the restricted experimental run, with only a slight decrease and increase in accuracy for $M_{\mathrm{HI}}$ and $M_{\mathrm{BH}}$, respectively. The results for our statistical validation are listed in Table~\ref{tab:table_2}, with the column under `True' corresponding to a machine learning model receiving true underlying $M_*$, SFR, and $Z$ values from \simba (coloured yellow in Figure~\ref{fig:histograms2}) and the column under `Hybrid' to predictions using the equilibrium model for these baryonic inputs while making use of merger tree information (coloured red in Figure~\ref{fig:histograms2}).

\begin{table}
 \caption{Statistical validation for the full experimental run including merger trees. The table lists the coefficient of determination ($R^2$) and Pearson's correlation coefficient ($\rho$) for different setups in alphabetically indicated columns. The column under `True' shows results for the prediction of neutral hydrogen ($M_{\mathrm{HI}}$), molecular hydrogen ($M_{\mathrm{H2}}$), and black hole mass ($M_{\mathrm{BH}}$) when feeding true underlying \simba target} values for stellar mass ($M_*$), star formation rate (SFR), and metallicity ($Z$) into the machine learning model. The column under `Hybrid' shows results for the prediction of the same properties as well as ${M_*, SFR, Z}$ when using the updated equilibrium model that includes merger tree information.
 \label{tab:table_2}
 \begin{tabular*}{\columnwidth}{@{}l@{\hspace*{30pt}}l@{\hspace*{30pt}}l@{\hspace*{30pt}}l@{\hspace*{30pt}}l@{}}
  \hline
  & \multicolumn{2}{@{}l}{\textbf{True}} & \multicolumn{2}{@{}l}{\textbf{Hybrid}}\\[2pt]
  \hline
  Variable & $R^2$ & $\rho$ & $R^2$ & $\rho$\\[2pt]
  \hline
  $M_{{\mathrm{{HI}}}}$ & $0.4873$ & $0.7110$ & $0.3650$ & $0.6533$\\[2pt]
  $M_{{\mathrm{{H2}}}}$ & $0.7826$ & $0.8863$ & $0.5470$ & $0.7513$\\[2pt]
  $M_{{\mathrm{{BH}}}}$ & $0.7774$ & $0.8907$ & $0.7143$ & $0.8815$\\[2pt]
  $M_*$ & $-$ & $-$ & $0.8229$ & $0.9478$\\[2pt]
  $\mathrm{{SFR}}$ & $-$ & $-$ & $0.7395$ & $0.8728$\\[2pt]
  $Z$ & $-$ & $-$ & $0.2167$ & $0.6641$\\[2pt]
  \hline
 \end{tabular*}
\end{table}

\section{Discussion}
\label{sec:discussion}

The hybrid nature of our approach, making use of both an established analytic formalism for galaxy evolution and machine learning, provides a number of advantages by combining, in an adage to timeless music, `the best of both worlds'. We derive a subset of baryonic parameters corresponding to dark matter haloes from the equilibrium model, an established formalism that we improve upon with a number of extensions. We include the gravitational free-fall time for mass accretion, as described in Section~\ref{sec:extension}, to further refine the model's capabilities of accurately tracing the evolution of haloes and their properties. In addition, we enable the equilibrium model to be fed complete largest-progenitor merger trees with corresponding redshift values, thus obviating the need to estimate halo masses at each time step to let the model follow the underlying mass evolution more closely. In adding these extension, one contribution of our work is the improvement of an established analytic formalism, making said formalism more suitable for fine-grained estimations and its application to N-body simulations and their merger trees.

Similarly, for the second half of our work, we merge the extended equilibrium model into a machine learning framework comprised of an ensemble of extra trees, as  discussed in Section~\ref{sec:framework}, as the latter has been shown to deliver the best results for the problem at hand~\citep{Ravanbaksh2017, Agarwal2018}. With increased dataset sizes in future research, which allows for larger training sets to fit machine learning models, we expect neural network models to catch up to, and surpass, the performance of tree-based ensembles. This can be realised with both larger-scale simulations and the combination of existing simulations, which we discuss further below. We contribute to the expanding literature on machine learning methods for cosmological simulations in general, and its application to, and analysis of, N-body and hydrodynamic simulations for baryonic property prediction in particular.

As these simulations are of crucial importance for several research areas in cosmology such as AGN feedback, survey analysis, covariance estimation, large-scale structure, and small-scale matter power behaviour, the resulting speed-up of going from N-body simulations to full hydrodynamic property sets per dark matter halo is of importance in the context of ever-growing simulation sizes and the analysis of upcoming surveys like Euclid and LSST~\citep{Racca2016, Ivezic2019}.

In a first step, we omit the inclusion of merger tree information and compare our hybrid approach, using the equilibrium model, to a hypothetical `perfect model', for which we simply use the true underlying properties normally derived from the equilibrium model as inputs. Given that the purpose of this part of the preliminary experimental run was to confirm the beneficial effect of including baryonic values in the input when predicting $M_{\mathrm{HI}}$, $M_{\mathrm{H2}}$, and $M_{\mathrm{BH}}$, the values for these properties fed into the machine learning model are the same as the target values, meaning that the model is given an unrealistic `ideal world' advantage. Even with perfect inputs for these three parameters, the model still performs slightly worse for $M_{\mathrm{HI}}$ when compared to $M_{\mathrm{H2}}$ and $M_{\mathrm{BH}}$, as can be seen in Table~\ref{tab:table_1}.

When compared to this hypothetical perfect scenario, a machine learning-only approach still performs reasonably well, confirming previous research on this topic. While we observe decreased accuracies for all assessed properties, the largest drop happens, again, for $M_{\mathrm{HI}}$, with the remaining parameters being less affected and $M_{\mathrm{BH}}$ experiencing the smallest decrease in accuracy, pointing towards a diminished reliance on baryonic inputs. Figure~\ref{fig:combined_predictions} also successfully recovers the $M_*-M_{\mathrm{BH}}$ relation, including the sharp drop at lower stellar masses, as a litmus test of the usability of our approach. The slight overprediction at higher stellar masses can be explained by more sparse data in the \simba dataset.

In the same preliminary experiment, we also establish that our hybrid approach underperforms, albeit still at a level useful for the completion of N-body simulations, when not making use of any baryonic inputs. This does, of course, not come as a surprise, as the preliminary experiment gauges initial halo masses based on a simple spline function fitted to the reduced equilibrium model itself, as shown in Figure~\ref{fig:splining}, and makes use of an invariant assumed initial redshift for this purpose. We also observe that the prediction of metallicity proves especially difficult in this setup.

We then incorporate the fully extended equilibrium model into our framework to test for the impact of including merger tree information. The results, depicted in Figure~\ref{fig:histograms2}, show that our approach recovers property distributions to a very reasonable degree, although we observe a slight underprediction of $M_*$ and overpredictions, especially of $M_{\mathrm{BH}}$ and $M_{\mathrm{HI}}$, which can be explained by the resolution challenge induced through a lower limit of $M_* \approx 10^{9.5} M_\odot$ in \simba. While distribution recoveries are a reliable way to ballpark the overall reproduction of values, it is important to statistically verify the results for a complete overview. The results in Table~\ref{tab:table_2} confirm that, for hypothetical perfect inputs of $M_*$, SFR, and $Z$, the results stay virtually the same for the assessed properties. Notably, the inclusion of the five last halo masses in the machine learning inputs does not improve the outcome, which further confirms that these properties do not rely on halo masses as much as the other inputs when predicted through a machine learning model.

For both $M_{\mathrm{HI}}$ and $M_{\mathrm{H2}}$, the results are slightly worse when compared to a machine learning-only approach as listed in Table~\ref{tab:table_1}, while the hybrid model outperforms on the prediction of $M_{\mathrm{BH}}$. Compared to the reduced equilibrium model without merger tree information, the predictions also outperform on both $M_{\mathrm{HI}}$ and $M_{\mathrm{BH}}$, while the accuracy for $M_{\mathrm{H2}}$ stays at the same level. For the baryonic properties, $M_*$, SFR, and $Z$, the extended setup clearly outperforms the reduced approach, with only the SFR remaining at a very similar, but satisfactory, accuracy. These results are especially useful as they not only demonstrate the viability of our hybrid approach, but also show that the use of baryon cycling parameters fitted on a reduced equilibrium model are viable on the extended version, confirming the robustness of the formalism.

While there is some degradation in the coefficient of determination, $R^2$, and Pearson's correlation coefficient, $\rho$, the degradation caused by using the equilibrium model as an intermediary is modest, for instance for HI with $\rho \approx 0.71$ in the ideal scenario with true baryonic inputs and $\rho=0.65$ when using the equilibrium model. HI is, in fact, the most difficult quantity to predict, while $\mbh$ $(\rho \approx 0.88)$ and $M_{H2}$ $(\rho \approx 0.75)$ are predicted with substantially higher fidelity. The equilibrium model also does a good job of reproducing the original \simba\ values of the input parameters $M_*$ and SFR, with $\rho \approx 0.95$ and $\rho \approx 0.87$, respectively, although the metallicity $Z$ is reproduced less effectively at $\rho \approx 0.66$. As the latter is predicted from the other inputs, as opposed to using the equilibrium model outputs, the accuracy is expectedly lower when compared to the other parameters, but still follows the intended distribution closely, as shown in the histograms in Figure~\ref{fig:histograms2}. Overall, this shows that the approach of using the equilibrium model to `pre-predict' a subset of baryonic quantities, which can be employed to improve training, is an effective approach towards more accurately bridging hydrodynamic simulations and large-volume N-body simulations.

For planned follow-up research, and aside from further investigations of the equilibrium model to improve the calculation of metallicities in the context of merger tree information, we propose the use of current developments in machine learning, especially in terms of boosting methods, active learning, and meta-ensembles, to alleviate this remaining issue. More generally, our work demonstrates that the hybrid model using analytic formalisms for baryonic properties outperforms in some areas, but underperforms in others. The use of such tailored meta-ensemble methods to base the weights placed on pure machine learning and hybrid approaches on the individual inputs could further improve the results, but would go beyond the scope of this work and is planned for future research. For an initial test, to evaluate the use of the equilibrium model against a pure machine learning approach, we provide the same type of extra trees ensemble with the entire set of redshift values and the evolutionary history of masses for each halo and find that such a replacement leads to little difference for the stellar mass ($\rho \approx 0.97$) and significantly underperforms for the star formation rate ($\rho \approx 0.50$), but outperforms on the metallicity ($rho \approx 0.79$) when compared to the results in Table~\ref{tab:table_2}. One possible explanation for this observation is that the metallicity can be better inferred by the machine learning ensemble due to shortcomings of a reduced equilibrium model for fitting baryonic cycling parameters. Conversely, the underperformance of pure machine learning in terms of the star formation rate can be due to the difficulty of approaching a perfect machine learning model for a given problem as well as due to the difficulty of dynamically modelling this property. For this reason, we plan on taking pure machine learning approaches into consideration specifically with regard to the problematic prediction of the metallicity, as well as for further investigations into correlations and informativeness in the context of machine learning models.

As with all research targeting specific datasets, both previous work on baryonic property prediction and this paper are limited to the recreation of specific simulations they are working with. Moreover, while machine learning excels at generalisation in terms of interpolation, meaning successful prediction within the value ranges presented by training datasets, extrapolation beyond these ranges is considerably more difficult and an active topic of research~\citep{Webb2020}. These issues could, for example, take the form of attempting to predict the described hybrid emulator to populate haloes from a larger box that happen to be more massive than the ones present in the \simba dataset used in our work, which presents a limitation of common machine learning approaches in general as well as our model in particular. Planned follow-up research will, for this reason, target the combination of data sources by assessing the compatibility of a variety of existing hydrodynamic simulations, allowing the framework to train on different pathways taken to model our Universe. Such a combination of simulations will also allow for larger training sets to fit more complex machine learning models, thus enabling the research community to revisit, for example, deep-layered neural network architectures for this purpose. Future improvements in cosmological simulations will further enhance the realism, which means that this framework can be reapplied and, thus, updated to develop into an increasingly robust predictive model.

In addition, sourcing simulations covering a larger range of values for the relevant features will help to alleviate the challenge of extrapolation mentioned above. For further follow-up research, we also suggest to assess the performance of hybrid approaches on haloes purposefully beyond the dataset ranges. Due to the modular nature of our two-part approach, both the machine learning component and the equilibrium model can also be replaced with novel developments in either of these areas in case they prove to be more suitable for the problem at hand. In this context, highly parallelised parameter estimation methods as described in Section~\ref{sec:introduction}, which draw from recent developments in both statistics and machine learning, will prove useful in potential follow-up research to constrain models relying on large datasets and likelihood calculations that require the running of computationally costly code. For this reason, we propose the use of these methods to investigate both the fast tuning of our model to given datasets and the application of such methods to simultaneous parameter optimisation for cosmological simulations.

An additional limitation is the narrower dynamic range in simulations, which means that the population of large-column simulations requires extrapolations beyond the dataset. As previously pointed out by~\citet{Agarwal2018}, this shortcoming could be tackled by using zoom simulations for dwarfs and galaxy clusters to retrieve anchor points for small and large halo masses~\citep{Cui2016}. Similarly, the focus on entire dark matter haloes using a largest-progenitor merger tree represents another limitation. The extended equilibrium model does, at this stage, not include satellite galaxies due to both data-side and model-side challenges. The former relies on the detection of subhaloes and numerical resolution in simulations, which often poses an issue, while the latter requires the inclusion of complete merger trees with all relevant progenitors in a suitable format, as well as the extension of the latter in the internal equilibrium model calculations.

\section{Conclusion}
\label{sec:conclusion}

In this paper, we introduce a hybrid framework using both machine learning and analytic components to predict baryonic galaxy properties based on dark matter halo information. In doing so, we lay the groundwork for a new class of merged approaches between analytic formalisms and machine learning. For this purpose, we extend the equilibrium model, a feedback-based description of the evolution of the stellar, gas, and metal content of galaxies, by including the ability to process largest-progenitor merger trees of dark matter haloes, as well as the free-fall time within haloes themselves. We then feed the partial baryonic outputs, together with dark matter properties, into a machine learning model that connects these properties to a full set of baryonic properties with stellar and black hole mass, neutral and molecular hydrogen, star formation rate, and metallicity, trained on the \simba\ cosmological hydrodynamic simulation. This framework is then able to predict with reasonable, though far from perfect, accuracy when compared to the true values taken from \simba.

We first introduce several modifications to the equilibrium model as described by \citet{Mitra2017}, including a slightly updated parameterisation of the baryon cycling parameters and the introduction of a delay time between halo and galaxy accretion given by the free-fall time. These minor updates improve the physical realism, but do not substantially change the goodness of fit versus observations. Next, we modify the equilibrium model to accept halo growth rates taken from merger trees, and use the equilibrium model to predict the baryonic properties $M_*$, SFR, and $Z$.

We feed this information, in addition to dark matter halo information, into an extremely randomised trees machine learning algorithm. The outputs of this process are various physical parameters that are not predicted directly by the equilibrium model. Here, we examine $\mbh$, $M_{HI}$, and $M_{H2}$, and train the extra trees model on \simba\ data. This now extends the predictive power of our framework to these additional quantities that are not directly predicted by the equilibrium model, using information from full hydrodynamic simulations. It is trivial to extend this to predicting other desired quantities, so long as they are outputs of a hydrodynamic simulation such as \simba\ on which they can be trained.

We test this approach by comparing two cases: In the first case, we input the true values for \{$M_*$, SFR, $Z$\} values from \simba, and then predict the remaining quantities; this is effectively the ideal case for machine learning predictions, since the extra baryonic inputs are taken directly from the hydrodynamic simulation itself. In the second case, we use the equilibrium model to obtain these properties from merger trees and associated redshifts, and then use those to predict the remaining quantities, namely $\mbh$, $M_{HI}$, and $M_{H2}$. Generally, we find that the second case has correlation coefficients that are not greatly degraded from the ideal case, indicating that the equilibrium model can provide a useful intermediary to improve baryonic property predictions within haloes at minimal computational cost, with the latter being a concern in modern cosmology for both temporal and environmental reasons. As with simulations in general, one thing that applies to this approach is that the framework learns to predict based on such a simulation, meaning that it learns not `true' physics but rather how properties are related in that simulation. Parameter inference thus also relates to how the simulation in question evolves galaxies, which should be taken into consideration when using our framework for such investigations.

In the future, we aim to extend this work in a three-pronged approach targeting all components of our framework: The equilibrium model is planned to include full merger trees with smaller progenitors, as well as satellite galaxies and black holes, to further push the model's accuracy and its predictive power for metallicities. On the machine learning side of our framework, we intend to make use of meta-learning approaches to weight input variables or the analytic and machine learning modules themselves. Lastly, advances in observational data and cosmological simulations allow for the equilibrium model to be fitted more accurately, and for the machine learning model to be trained on a wide variety of simulation approaches. As the equilibrium model provides reliable Bayesian posteriors, planned follow-up research will also investigate constraints on dark matter.

In addition, we plan to investigate modelling the correlated scatter in galaxy quantities more accurately. In particular, the division between quenched and star-forming galaxies, as well as the associated trends in gas content and other properties, has often been challenging to recover using machine learning.  These issues suggest that perhaps a combination of methodologies including both classification and regression may be more optimal. Alternatively, different machine learning approaches such as generative adversarial networks may be more effective at picking out the more subtle trends in the galaxy population.

The resulting framework will, in principle, have a wide applicability for both cosmology and galaxy evolution studies, including populating dark matter-only simulations, examining the physical constraints on baryon cycling parameters, and investigating environmental trends in the galaxy population such as assembly bias. Machine learning applied to galaxy evolution is still a developing field, but offers great promise for delivering the most accurate mock universes incorporating information from both high-resolution hydrodynamic simulations and large-volume N-body simulations.

\section*{Acknowledgements}

We thank Joe Zuntz for helpful discussions. BM acknowledges the support of a Principal's Career Development Scholarship from the University of Edinburgh. RD acknowledges support from Wolfson Research Merit Award WM160051 from the U.K. Royal Society. WC acknowledges the support from the European Research Council under grant 670193. The \simba\ simulation was run on the DiRAC@Durham facility managed by the Institute for Computational Cosmology on behalf of the STFC DiRAC HPC Facility. The equipment was funded by BEIS capital funding via STFC capital grants ST/P002293/1, ST/R002371/1 and ST/S002502/1, Durham University and STFC operations grant ST/R000832/1. DiRAC is part of the National e-Infrastructure.

\section*{Data Availability}

The data underlying this article were accessed from the \simba Project Repository (\href{http://simba.roe.ac.uk/}{simba.roe.ac.uk}). The derived data generated in this research will be shared on reasonable request to the corresponding author.

\bibliographystyle{mnras}
\bibliography{references}

\bsp	
\label{lastpage}
\end{document}